\documentclass [twocolumn,aps,prb,noshowpacs]{revtex4-1}
\usepackage{epsfig,graphicx,amssymb,color,amssymb,amsmath,mathrsfs,bm}
\pdfoutput=1

\begin{document}

\title{Second-harmonic generation in nonlinear plasmonic lattices enhanced by quantum emitter gain medium}

\author{Maxim Sukharev}
\email[]{corresponding author: maxim.sukharev@asu.edu}
\affiliation{Department of Physics, Arizona State University, Tempe, AZ 85287, USA }
\affiliation{College of Integrative Sciences and Arts, Arizona State University, Mesa, AZ 85201, USA}

\author{Oleksiy Roslyak}
\affiliation{Physics and Engineering Physics, Fordham University, Bronx, NY 10458, USA}

\author{Andrei Piryatinski}
\email[]{apiryat@lanl.gov}
\affiliation{Theoretical Division, Los Alamos National Laboratory, Los Alamos, NM 87545, USA }

\begin{abstract}
We report on theoretical study of second-harmonic generation (SHG) in plasmonic nanostructures interacting with two-level quantum emitters (QE) under incoherent energy pump. We generalize driven-dissipative Tavis-Cummings model by introducing anharmonic surface plasmon-polariton (SPP) mode coupled to QEs and examine physical properties of corresponding SPP-QE polariton states. Our calculations of the SHG efficiency for strong QE-SPP coupling demonstrate orders of magnitude enhancement facilitated by the polariton gain. We further discuss time-domain numerical simulations of SHG in a square lattice comprised of Ag nanopillars coupled to QEs utilizing fully vectorial nonperturbative nonlinear hydrodynamic model for conduction electrons coupled to Maxwell-Bloch equations for QEs. The simulations support the idea of gain enhanced SHG and show orders of magnitude increase in the SHG efficiency as the QEs are tuned in resonance with the lattice plasmon mode and brought above the population inversion threshold by incoherent pump. By varying pump frequency and tuning QEs to a localized plasmon mode, we demonstrate further enhancement of the SHG efficiency facilitated by strong local electric fields. The incident light polarization dependence of the SHG is examined and related to the symmetries of participating plasmon modes.
\end{abstract}

\date{\today}
 
\maketitle


\section{Introduction}

Research in optics of nanomaterials including the nanoplasmonics has significantly advanced due to both tremendous progress in nanofabrication and further characterization employing various new ultrafast spectroscopic techniques.\cite{StockmanOptEx:2011} The nanoplasmonics has enjoyed substantial growth scrutinizing physical properties of propagating \cite{WeinerRepProgPhys:2009,GarciaVidalRMP:2010} and localized surface plasmon-polaritons (SPP) and utilizing their unique polarization features to control light propagation \cite{SukharevNanoLet:2006} at various metal-dielectric interfaces.\cite{EbbesenPhysToday:2008} The nonlinear optics of plasmonic nanosctructures has quickly become a hot research field.\cite{KauranenNatPhot:2012} Applications range from optical bistability \cite{WurtzPRL:2006} through second-harmonic generation (SHG)\cite{PanoiuJ.Opt:2018} to difference frequency generation.\cite{SiderisOptLett:2019}  For instance, by manipulating geometrical parameters one can achieve substantial increase in four-wave mixing processes.\cite{BlechmanNanoLett:2018} Both experiment and theory show that the SHG by plasmonic nanoparticles with no center of inversion symmetry is extremely sensitive to particle shape even within few nanometers opening new opportunities to designing functional materials for frequency conversion applications.\cite{MaekawaLPCC:2020} 

On the other hand, strong spatial localization of SPPs could be used to investigate how ensembles of quantum emitters (QEs), e.g., organic dyes or semiconductor quantum dots, behave if resonantly coupled to corresponding surface electromagnetic modes.\cite{TormaRepPrgoPhys:2014}  When in such systems the coupling strength surpasses all damping rates, the so-called strong coupling regime is achieved resulting in formation of QE-SPP hybridize polariton states \cite{SukharevJPCM:2017}. These states will be referred to as the polaritons throughout this paper. Recent studies show that in the strong coupling regime the second-order nonlinear processes that are mostly caused by highly polarizable metal electrons are significantly altered by QEs.\cite{DrobnyhJCP:2020} The polariton states contribute on equal footing to the SHG resulting in detectable Rabi splitting of the signal even at a single nanoparticle level.\cite{LiarXiv:2020} Numerical simulations show that the presence of a gain medium in core-shell metal-dielectric nanostructures supporting localized surface-plasmon modes, leads to an enhancement of SHG.\cite{PanSciRe:2017} However, mechanisms of enhanced SHG by the polariton states contributed by QEs in the population inversion regime require theoretical examination.           

From quantum plasmonics point of view, highly polarizable metal nanostructures can play a role of open optical cavities.\cite{TameNatPhys:2013}  Corresponding plasmonic cavity modes are typically described as perfect harmonic oscillators while the nonlinearity arises from the interaction with QEs. In strong coupling regime such a nonlinear system shows unique properties of the polariton states. As recently demonstrated for plasmonic lattices, these states are responsible for nonlinear emission, lasing~\cite{RamezaniPRL:2018,RichterPRB:2015,ZhangJCP:2015,HoangNL:2017,RamezaniOptica:2017,RamirezAdvMat:2019,Guan_ASCnano:2020,Pusch_ACSnano:2012,Winkler_ACSnano:2020}, non-equilibrium superradiant phase transitions~\cite{Piryatinski_PRR:2020}, and Bose-Einstein condensation~\cite{RamezaniOptica:2017,MartikainenPRA:2014,ZasterJPCS:2016,HakalaNP:2018,RamezaniJOSAB:2019}. An ability of plasmonic nanostructures to support anharmonic response~\cite{Butet_ACS_Nano:2015} brings an interesting notion of anharmonic quantum plasmonic cavities. In this paper, we study an interplay of the QE nonlinearity and anharmonicity effects on polariton state structure and nonlinear optical response in such plasmonic cavities. 

For this purpose, we generalize the driven dissipative Tavis-Cummings model allowing for lasing phase transitions~\cite{Kirton_AdvQT:2019,RichterPRB:2015,Piryatinski_PRR:2020}. Specifically, we introduce an ensemble of two-level QEs subject to incoherent energy pump and dissipation which interacts with a single anharmonic SPP mode. In practice, incoherent pumping can be realized via optical excitation of high electronic states of QEs followed by the excitation relaxation to the optically active states or via direct carrier injection. If the pumping rate is high enough to create QE population inversion, a gain builds up compensating plasmonic losses. Thus, we focus on examining formation of the polariton gain and subsequent enhancement of SHG in the strong QE-SPP coupling regime for various strengths of SPP anharmonicity. 

Although we quantize the anharmonic SPP mode, our results are valid in the semi-classical limit holding for high population of SPP quanta and typical for SHG experiment. To this extent our minimal model is capable of providing an upper limit estimate for the SHG efficiency. To account for the effect of complex structure of  SPP modes in plasmonic lattices, we further performed time-domain numerical simulations of SHG in 2D lattice of Ag nanopillars coupled to QEs. These calculations utilize fully vectorial nonlinear hydrodynamic model for conduction electrons coupled to Maxwell-Bloch equations describing QEs. The simulation results can be qualitatively interpreted with the help of our minimal model and give a lower boundary for experimentally expected values of SHG efficiency. On the other hand the simulations reveal enhanced SHG dependence on the incident light polarization which can be experimentally verified.       

The paper is organized as follows. In Sec.~\ref{Sec:anhTCM}, we discusses the anharmonic Tavis-Cummings model and its mean-field steady states pinpointing the physics of SHG enhancement by polariton gain. In Sec.~\ref{Sec:MaxSim}, we discuss the results of the SHG simulations in Ag nanopillar lattice identifying  role of various SPP modes. Conclusions are drawn in Sec.~\ref{Sec:Concl}.


\section{Driven-dissipative anharmonic Tavis-Cummings model}
\label{Sec:anhTCM}

\begin{figure}[b]
\begin{center}
\epsfig{file=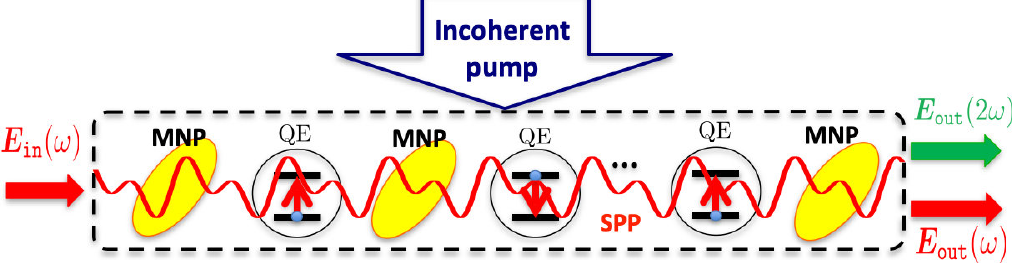,width=3.2in}
\end{center}
\caption{Schematics of adopted driven-dissipative anharmonic Tavis-Cummings model.}
\label{Fig:Model}
\end{figure}

Our model illustrated in Fig.~\ref{Fig:Model} consists of an array of metal nanoparticles (MNPs) supporting delocalized SPP mode treated as an anharmonic oscillator described by a bosonic operator $\hat\psi$ with hermitian conjugate $\hat\psi^\dag$. The SPP interacts with an ensemble of ${\cal N}_o$ identical two-level QEs described by spin operators $\hat s_n^\pm=\hat s_n^x \pm i\hat s^y_n$, $s_n^z$, expressed in terms of the Pauli SU(2) operators as $\hat s_n^{j}=\frac{1}{2}\hat\sigma_n^{j}$ with $j=x,y,z$ and site index $n=\overline{1,{\cal N}_o}$. The QEs are subject to incoherent energy pump whereas the SPP interacts coherently with an incident monochromatic laser field $\bm E_\texttt{in}(\omega)$. By taking into account that the SPP transition dipole, $\mu_{sp}$, exceeds the QE transition dipole by orders of magnitude, we consider the output electric field generated by the SPP only. The field is further partitioned into two modes $\bm E_\texttt{out}(\omega)\sim\bm \mu_{sp}\hat\psi(\omega)$  and  $\bm E_\texttt{out}(2\omega)\sim\bm \mu_{sp}\hat\psi(2\omega)$  oscillating with fundamental and second-harmonic frequencies, respectively.  Both QEs and MNPs interact with the environment causing dissipative processes. 

The coherent dynamics of our model is described by the Hamiltonian 
\begin{eqnarray}
\label{Ham-def}
\hat {\cal H} &=&  \omega_o\hat\psi^\dag\hat\psi 
+\alpha \left(\hat\psi^\dag \hat\psi^\dag\hat\psi+\hat\psi^\dag\hat\psi \hat\psi\right)
\\\nonumber &+&
\omega_o\left(\sum\limits_{n=1}^{{\cal N}_o}\hat s_n^z+\frac{{\cal N}_o}{2}\right) 
+\lambda \sum\limits_{n=1}^{{\cal N}_o}\left(\hat\psi^\dag \hat s_n^{-} + \hat s_n^{+} \hat\psi \right)
\\\nonumber &-&		
	\Omega_{\texttt{in}}\left(\hat\psi e^{i\omega t} + \hat\psi^\dag e^{-i\omega t}\right),
\end{eqnarray}
represented in units of $\hbar$. Here, we set the SPP and all QEs to have the same transition frequency $\omega_o$ and identical coupling rate $\lambda$. The second term of the Hamiltonian, reflects the SPP cubic anharmonicity with the coupling rate $\alpha$. Interaction between the SPP and incident driving field, $ \bm E_\texttt{in}$, is accounted for in the last term of the Hamiltonian where the coupling strength is represented by the Rabi frequency $\Omega_\texttt{in}=(\bm \mu_{sp}\cdot \bm E_\texttt{in})/\hbar$. 

To include the driven-dissipative dynamics, we employ Heisenberg equation of motion in the form
\begin{eqnarray}
\label{HsEqMo-def}
\partial_t \hat{\cal O} &=& i\left[\hat{\cal H}, \hat{\cal O} \right]
        +2\gamma_{sp} \hat{\cal D}_{\hat \psi}[\hat {\cal O}]
	+\gamma_\uparrow\sum\limits_{n=1}^{{\cal N}_o} \hat{\cal D}_{\hat s^{+}_n}[\hat {\cal O}]
\\\nonumber &+&	
	\gamma_\downarrow\sum\limits_{n=1}^{{\cal N}_o}\hat{\cal D}_{\hat s^{-}_n}[\hat {\cal O}]
	+ \gamma_\phi\sum\limits_{n=1}^{{\cal N}_o}\hat{\cal D}_{\hat s_n^z}[\hat {\cal O}],
\end{eqnarray}
where the first term in the right-hand side is responsible for coherent dynamics due to the Hamiltonian $\hat {\cal H}$.  The rest of the terms describe the SPP decay with the rate $2\gamma_{sp}$, incoherent pumping (population decay) of QEs with the rate $\gamma_\uparrow$ ($\gamma_\downarrow$), and QE pure dephasing with the rate $\gamma_\phi$, respectively. Associated Lindblad operator in the Heisenberg representation reads
\begin{eqnarray}
\label{LdEqMo-def}
\hat{\cal D}_{\hat C}[\hat {\cal O}] &=& \frac{1}{2}\hat C^\dag\left[ \hat {\cal O},  \hat C \right] 
				+\frac{1}{2}\left[ \hat C^\dag ,\hat {\cal O}\right]  \hat C.
\end{eqnarray}

In the limit of $\alpha=\Omega_\texttt{in}=0$, Eqs.~\eqref{Ham-def}--\eqref{LdEqMo-def}  recover open Tavis-Cummings model allowing for the lasing phase transition \cite{Kirton_AdvQT:2019,RichterPRB:2015,Piryatinski_PRR:2020}. Our generalization of the model includes the SPP anharmonicity and external coherent laser field. As demonstrated below this generalization constitutes a minimal model to examine SHG and effect of the QE induced gain to enhance this process.

 \subsection{Mean-field steady states}
 
Starting with Eqs.~\eqref{HsEqMo-def} and \eqref{LdEqMo-def} and the  Hamiltonian~\eqref{Ham-def}, we derive the Heisenberg operator equations for $\hat{\cal O} = \{\hat\psi, \hat s_n^{-}, \hat s_n^z\}$ and adopting the mean-field approximation replace the operators with their averages. The result is
\begin{eqnarray}
\label{MF-psi}
\partial_t \varphi &=& -(i\omega_o +\gamma_{sp})\varphi-i\sqrt{{\cal N}_o}\alpha\left( \varphi^2 +2|\varphi|^2 \right)
\\\nonumber
&~& -i\sqrt{{\cal N}_o}\lambda  s_{-} -\frac{i\Omega_\texttt{in}}{\sqrt{{\cal N}_o}}e^{-i\omega t }, 
\\\label{MF-sm}
\partial_t s_{-}&=&-\left(i\omega_o+\gamma_o\right) s_{-}+2i\sqrt{{\cal N}_o}\lambda s_z \varphi,
\\\label{MF-sz}
 \partial_t s_z &=& -\gamma_t\left(s_z-\frac{d_o}{2}\right) + i\sqrt{{\cal N}_o}\lambda \left(\varphi^* s_{-}- s_{+}\varphi\right), 		
\end{eqnarray}
where we introduced normalized SPP, $\varphi = \langle \hat\psi\rangle/\sqrt{{\cal N}_o}$ and QE, $s_\pm = \sum\limits_{n=1}^{{\cal N}_o} \langle \hat s_n^{\pm}\rangle/{\cal N}_o$, coherences, and normalized QE population $s_z= \sum\limits_{n=1}^{{\cal N}_o} \langle \hat s_n^z\rangle/{\cal N}_o$  varying in the interval $[-1/2,1/2]$. Note that the effective QE-SPP coupling, $\sqrt{{\cal N}_o}\lambda$, and anharmonicity,  $\sqrt{{\cal N}_o}\alpha$, rates scale with the system size as $\sqrt{{\cal N}_o}$ and can be enhanced by increasing number of QEs. 

The SPP dephasing rate, $\gamma_{sp}$, entering Eq.~\eqref{MF-psi} is the same as in Eq.~\eqref{HsEqMo-def}. However, the QE dissipation rates in Eqs.~\eqref{MF-sm} and \eqref{MF-sz} become linear combinations of those in Eq.~\eqref{HsEqMo-def}. Specifically, the QE dephasing and population decay rates are $\gamma_o=\gamma_\downarrow/2+\gamma_\uparrow/2+\gamma_\phi$ and  $\gamma_t=\gamma_\uparrow+\gamma_\downarrow$, respectively. Notice their dependence on the incoherent pumping rate, $\gamma_{\uparrow}$. Furthermore, Eq.~\eqref{MF-sz} contains the quantity 
\begin{eqnarray}
\label{do-def}
d_o=\frac{\gamma_\uparrow-\gamma_\downarrow}{\gamma_\uparrow+\gamma_\downarrow},
\end{eqnarray}
referred to as the population inversion parameter. Depending on the incoherent pumping rate, $\gamma_\uparrow$, it can be $-1<d_o<0$ indicating that $\gamma_\uparrow<\gamma_\downarrow$ and QEs are pumped below population inversion threshold. Increase in the pumping to $\gamma_\uparrow>\gamma_\downarrow$ brings QEs to the inversion regime with $0<d_o<1$.

To solve Eqs.~\eqref{MF-psi}--\eqref{MF-sz} accounting for the fundamental frequency and the second-harmonic response, we transform the coherences into rotating frames as
\begin{eqnarray}
\label{psi1-psi2-def}
&~&\varphi(t) = \tilde\varphi_1(t) e^{-i\omega_L t} + \tilde\varphi_2(t) e^{-2i\omega_L t},
\\\label{sm-swa-def}
&~&s_{-}(t) = \tilde s_{-}(t) e^{-i\omega_L t},
\end{eqnarray}
where slowly varying amplitudes for the fundamental SPP, $\tilde\varphi_1(t)$, and QE, $\tilde s_{-}(t)$, coherences as well as for the second-harmonic SPP coherence $\tilde\varphi_2(t)$ are introduced. Substitution of Eqs.~\eqref{psi1-psi2-def} and \eqref{sm-swa-def} into Eqs.~\eqref{MF-psi}--\eqref{MF-sz} and subsequent elimination of the fast oscillating terms results in the following set of coupled equations of motion for the slowly varying amplitudes
\begin{eqnarray}
\label{psi1-eqm-swa}
\partial_t \tilde\varphi_1 &=& -\left(i\delta\omega +\gamma_{sp}\right)\tilde\varphi_1-2i\alpha \tilde\varphi_1^{*}\tilde\varphi_2
\\\nonumber &~&
- i\sqrt{{\cal N}_o} \lambda  \tilde s_{-} +\frac{i\Omega_\texttt{in}}{\sqrt{{\cal N}_o}}e^{-i(\omega-\omega_o+\delta\omega)t },
\\\label{psi2-eqm-swa}
\partial_t \tilde\varphi_2 &=& - \left(i[2\delta\omega-\omega_o] +\gamma_{sp}\right)\tilde\varphi_2- i\sqrt{{\cal N}_o}\alpha \tilde\varphi_1^2
\\\label{sm-eqm-swa}
\partial_t \tilde s_{-}&=&-\left(i\delta\omega+\gamma_o\right)\tilde s_{-}+2i\sqrt{{\cal N}_o}\lambda s_z \tilde\varphi_1,
\\\label{sz-eqm-swa}
 \partial_t s_z &=& -\gamma_t\left(s_z-\frac{d_o}{2}\right) + i\sqrt{{\cal N}_o}\lambda \left(\tilde\varphi_1^* \tilde s_{-}- \tilde s_{+} \tilde\varphi_1\right).\;\;\;		
\end{eqnarray}
Here, $\delta\omega = \omega_o-\omega_L$ is the frequency detuning parameter. 

Now, let us examine the steady states of Eqs.~\eqref{psi1-eqm-swa}--\eqref{sz-eqm-swa} focusing on the properties of fundamental SPP coherence, $\tilde\varphi_1$. Keeping in mind that the external field is weak enough to have perturbative effect on the steady states, we set $\Omega_\texttt{in}=0$. By solving Eqs.~\eqref{psi2-eqm-swa}--\eqref{sz-eqm-swa} for the steady states and using these solutions to eliminate the second-harmonic coherence and the QE variables in Eq.~\eqref{psi1-eqm-swa}, we obtain the following nonlinear algebraic equation,  
\begin{eqnarray}
\label{psi1-stst-eq}
\Sigma\left\{|\tilde\varphi_1|\right\}\tilde\varphi_1 =0,
\end{eqnarray}
determining the steady states of $\tilde\varphi_1$. Here, the prefactor 
\begin{eqnarray}
\label{Sigma-def}
\Sigma\left\{|\tilde\varphi_1|\right\} =  i\Delta\Omega\left\{|\tilde\varphi_1|\right\} +\Gamma\left\{|\tilde\varphi_1|\right\}
 -G\left\{|\tilde\varphi_1|\right\}, 	
\end{eqnarray}
contains the real and imaginary parts functions of $|\tilde\varphi_1|$. The imaginary part   
\begin{eqnarray}
\label{Omega-phi-def}
\Delta\Omega\left\{|\tilde\varphi_1|\right\}&=& \delta\omega
-  \frac{2{\cal N}_o\alpha^2(2\delta\omega-\omega_o) }
		{(2\delta\omega-\omega_o)^2 +\gamma_{sp}^2} |\tilde\varphi_1|^2
\\\nonumber &+&		
		\frac{d_o{\cal N}_o \lambda^2 \delta\omega}{\delta\omega^2+\gamma_o^2+4{\cal N}_o\lambda^2[\gamma_t/\gamma_o] |\tilde\varphi_1|^2},	
\end{eqnarray}
describes renormalization of the detuning parameter by self-energies due to the anharmonic interaction (second term) and the SPP-QE coupling (third term). 

The real part of Eq.~\eqref{Sigma-def} describes an interplay of two dephasing terms, specifically,  
\begin{eqnarray}
\label{Gmma-phi-def}
\Gamma\left\{|\tilde\varphi_1|\right\}&=&\gamma_{sp}  + \frac{2{\cal N}_o \alpha^2 \gamma_{sp} |\tilde\varphi_1|^2 }
		{(2\delta\omega-\omega_o)^2 +\gamma_{sp}^2},
\end{eqnarray}
being the SPP dephasing due to the energy decay into heat (first term) and due to the energy transfer to the second-harmonic (second term). The other contribution to the real part is the gain rate
\begin{eqnarray}
\label{Gain-def}
G\left\{|\tilde\varphi_1|\right\}&=& \frac{d_o{\cal N}_o \lambda^2 \gamma_o}{\delta\omega^2+\gamma_o^2+4{\cal N}_o\lambda^2[\gamma_t/\gamma_o] |\tilde\varphi_1|^2},
\end{eqnarray}
compensating the dephasing term provided $0<d_o \leq 1$, i.e., inverted QE population. Below the population inversion threshold, $-1\leq d_o<0$, sign of Eq.~\eqref{Gain-def} becomes negative resulting in additional contribution to the losses.

Equation~\eqref{psi1-stst-eq} has a trivial solution $\tilde\varphi_1=0$ and a nontrivial solution defined by $\Sigma\left\{|\tilde\varphi_1|\right\}=0$. The latter satisfies if the real part (frequency shift) and the imaginary part (gain-loss term) of Eq.~\eqref{Sigma-def} both become zero, i.e., 
\begin{eqnarray}
\label{dOmg}
&~&\Delta\Omega\left\{|\tilde\varphi_1|\right\} = 0,
\\\label{dGmma}
&~&\Gamma\left\{|\tilde\varphi_1|\right\}-G\left\{|\tilde\varphi_1|\right\} =0.
 \end{eqnarray}
These conditions determine the lasing phase transition in the driven-dissipative Tavis-Cummings model. Roots of Eqs.~\eqref{dOmg} and \eqref{dGmma} are coherences $|\tilde\varphi_1|$ spontaneously appearing above critical coupling,  $\lambda_c$, and the detuning parameter $\delta\omega$ stands for the lasing frequency shift.~\cite{Kirton_AdvQT:2019,RichterPRB:2015,Piryatinski_PRR:2020} Below $\lambda_c$, the trivial solution, $\tilde\varphi_1=\delta\omega=0$ holds.

\subsection{Polariton dispersion and SHG efficiency}
\label{Sec:DispEffc}

Having defined the steady state solutions, we proceed to examine both linear and non-linear responses induced by the incident driving field. For the linear response, we use time-dependant equation of motion for the SPP coherence at the fundamental frequency
\begin{eqnarray}
\label{psi1-rdcd}
\left(\partial_t + \Sigma\left\{|\tilde\varphi_1|\right\} \right)\tilde\varphi_1 
=\frac{i\Omega_\texttt{in}}{\sqrt{{\cal N}_o}}e^{-i(\omega -\omega_o+\delta\omega)t }, 
\end{eqnarray}
derived in the same way as Eqs.~\eqref{psi1-stst-eq}--\eqref{Gain-def} but retaining the time-dependence. We further seek solution of this equation in terms of time-dependant perturbations $\delta\varphi_1(t)$ of the mean-field steady state $\bar\varphi_1$. Making the substitution of  $\tilde\varphi_1(t)=\bar\varphi_1+\delta\varphi_1(t)$ into Eq.~\eqref{psi1-rdcd} and complementing the resulting equation with complex conjugate, we derive linearized  equations of motion in the matrix form
\begin{eqnarray}
\label{dphi1-eqm}
\left(\mathbb{I}\partial_t + {\cal M}\right)\delta\Phi = \frac{i\Omega_\texttt{in}}{\sqrt{{\cal N}_o}} {\cal F}(t),
\end{eqnarray}
with the column vectors $\delta\Phi = (\delta\varphi_1,\delta\varphi^*_1)^\texttt{T}$ and ${\cal F}(t)=\left(e^{-i(\omega-\omega_o+\delta\omega)t},-e^{i(\omega-\omega_o+\delta\omega)t}\right)^\texttt{T}$. Here, $\mathbb{I}$ denotes $2\times 2$ unit matrix and 
\begin{eqnarray}
\label{Mstb}
{\cal M} =
   \begin{pmatrix} 
     {\cal M}_\texttt{d}   &  {\cal M}_\texttt{o}   \\
     {\cal M}_\texttt{o} ^{*}  &  {\cal M}_\texttt{d}^{*}, \\
   \end{pmatrix},
 \end{eqnarray}
stands for the so-called stability the matrix with matrix elements 
\begin{eqnarray}
\label{Md-def}
{\cal M}_\texttt{d}&=& \Sigma\left\{|\bar\varphi_1|\right\} + \bar\varphi_1\partial_{\bar\varphi_1}\Sigma\left\{|\bar\varphi_1|\right\},
 \\\label{Md-def}
{\cal M}_\texttt{o}&=&  \bar\varphi_1\partial_{\bar\varphi^*_1}\Sigma\left\{|\bar\varphi_1|\right\}, 
 \end{eqnarray}
depending on $\Sigma\left\{|\bar\varphi_1|\right\}$ (Eqs.~\eqref{Sigma-def}--\eqref{Gain-def}) and its functional derivative both functions of the steady state solutions, $\bar\varphi_1$, of Eq.~\eqref{psi1-stst-eq}. 

Solution of Eq.~\eqref{dphi1-eqm} in the presence of oscillating incident field defines SPP polarization at the fundamental frequency
\begin{eqnarray}
\label{dphi_linsl}
\mu_{sp}\delta\varphi_1(t)&=&\chi^{(1)}(\omega)E_\texttt{in}
e^{-i(\omega-\omega_o+\delta\omega)t },
\end{eqnarray}
where the linear susceptibility is
\begin{eqnarray}
\label{chi1-def}
\chi^{(1)}(\omega_L)&=&\frac{\mu^2_{sp}}{{\cal N}_o\hbar}
\left[\frac{u_+^Lu_+^R}{\omega-\omega_o+\delta\omega+i\Lambda_+}   
\right.\\\nonumber &+& \left.
\frac{u_-^Lu_-^R}{\omega-\omega_o+\delta\omega+i\Lambda_-}
\right],
\end{eqnarray}
and depends on the stability matrix, ${\cal M}$, eigenvalues 
\begin{eqnarray}
\label{Lmbd_pm-def}
\Lambda_\pm = \text{Re}\left[{\cal M}_d\right] \pm\sqrt{\left|{\cal M}_o
\right|^2-\text{Im}\left[{\cal M}_d\right] ^2 }
\end{eqnarray}
and the upper components of orthonormal left $\delta\Phi^{L}_\pm=(u^{L}_\pm,v^L_\pm)$ and right  $\delta\Phi^{R}_\pm=(u^{R}_\pm,v^R_\pm)^\texttt{T}$ eigenvectors. Poles of the linear susceptibility 
\begin{eqnarray}
\label{plrt-poles}
\zeta_\pm=\omega_o-\delta\omega-i\Lambda_\pm,
\end{eqnarray}
determine the polariton frequency, $\omega_\pm=\text{Re}\left[\zeta_\pm \right]$, and dephasing rate $\gamma_\pm=\text{Im}\left[\zeta_\pm \right]$. 

Below the critical coupling $\lambda<\lambda_c$, $\tilde\varphi_1=\delta\omega=0$ simplifying  Eq.~\eqref{Lmbd_pm-def} to the form
\begin{eqnarray}
\label{Lmbd_pm-triv}
\Lambda_\pm = \gamma_{sp}-d_o\frac{{\cal N}_o \lambda^2 }{\gamma_o},
\end{eqnarray}
where the second term in the right-hand side is an explicit representation of the polariton gain term $G\{|\bar\varphi_1|=0\}$ (Eqs.~\eqref{Gain-def}). According to Eqs.~\eqref{plrt-poles} and \eqref{Lmbd_pm-triv}, the polariton frequency  is constant $\omega_\pm=\omega_o$. In contrast, the polariton dephasing $\gamma_\pm =\Lambda_\pm$, decreases with the coupling increase until the gain rate fully compensates the losses, i.e., $\gamma_\pm=0$. This condition determines the critical coupling 
\begin{eqnarray}
\label{lmbd_c-def}
{\cal N}_o\lambda^2_c = \frac{\gamma_{sp}\gamma_o}{d_o},
\end{eqnarray}
where $\gamma_o$ depends on the incoherent pumping rate as well as $d_o$. Equation~\eqref{lmbd_c-def} is know as the lasing threshold in open Tavis-Cummings model~\cite{RichterPRB:2015,Piryatinski_PRR:2020} which in our generalized case does not acquire dependence on the anharmonicity rate, $\alpha$. According to Eq.~\eqref{lmbd_c-def}, requirement that $\lambda>\lambda_c$ implies that ${\cal N_o}\lambda^2 > \gamma_{sp}\gamma_o$, i.e., the QE-SPP coupling strength exceeds the system losses. This is a definition of the strong coupling regime.\cite{RamezaniJOSAB:2019}

In the strong coupling regime, $\lambda>\lambda_c$, the polariton splitting is determined by the square root expression in Eq.~\eqref{Lmbd_pm-def} containing $\left|{\cal M}_o\right|^2-\text{Im}\left[{\cal M}_d\right] ^2 $. Depending on the incoherent pumping rate and anharmonicity values this difference can become either positive or negative. In the former case, the square root becomes real and the polariton dephasing $\gamma_\pm$ experiences splitting into two branches while the frequency $\omega_\pm$ stays degenerate. In the latter case, the square root becomes imaginary leading to splitting of the polariton energy $\omega_\pm$ but the dephasing $\gamma_\pm$ stays degenerate. As shown below, the former situation represents lasing/amplification regime whereas the later one does not.   

\begin{figure*}[t]
\begin{center}
\epsfig{file=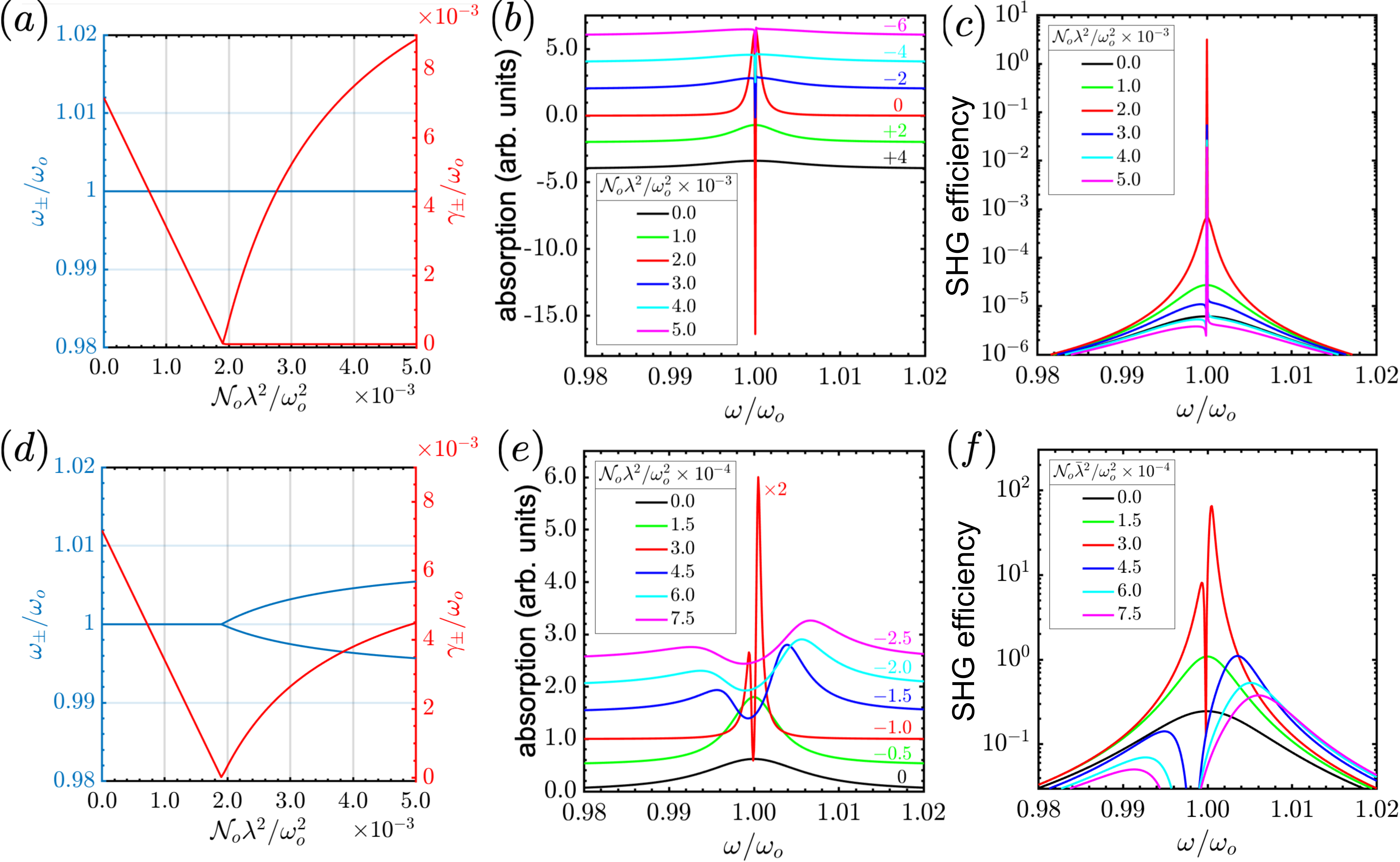,width=5.5in}
\end{center}
\caption{(a,d) Polariton frequency, $\omega_\pm$ and dephasing rate $\gamma_\pm$ dispersion entering strong coupling regime above critical coupling at ${\cal N}_o\lambda_c^2/\omega_o^2=1.9\times 10^{-3}$ (b,e) Linear absorption/emission spectra of the polariton states, (c,f) SHG efficiency (Eq.~\eqref{I2toI1}) calculated for various QE-SPP coupling strengths ${\cal N}_o\lambda^2/\omega_o^2$. Panels (a)-(c) present weak anharmonic coupling, $\sqrt{{\cal N}_o}\alpha/\omega_o=10^{-3}$, and (d)-(f) strong one $\sqrt{{\cal N}_o}\alpha/\omega_o=0.2$. Curves in panels (b) and (e) are vertically staggered for clarity with the color coded numbers indicting associated base shift values.} 
\label{Fig:Plr}
\end{figure*}

Assuming that the anharmonicity rate $\alpha\ll\omega_o,\lambda$, we perturbatively calculate the second-harmonic response, $\delta\varphi_2(t)$,  using Eq.~\eqref{psi2-eqm-swa} with the linear solution (Eqs.~\eqref{dphi_linsl} and \eqref{chi1-def}), $\delta\varphi_1^2$,   as the driving force. Accordingly, solution of Eq.~\eqref{psi2-eqm-swa} gives  the following SPP second-harmonic polarization  
\begin{eqnarray}
\label{dphi2-sol}
\mu_{sp}\delta\varphi_2(t)&=&\chi^{(2)}(2\omega,\omega,\omega)E^2_\texttt{in}
e^{-2i(\omega-\omega_o+\delta\omega)t },
\end{eqnarray}
in terms of the second-order nonlinear susceptibility 
\begin{eqnarray}
\label{chi2-def}
\chi^{(2)}(2\omega,\omega,\omega)&=&\frac{\sqrt{{\cal N}_o}\alpha \left[\chi^{(1)}(\omega)\right]^2}{\omega_o-2\omega-i\gamma_{sp}},
\end{eqnarray}
which in turn depends on $\chi^{(1)}(\omega)$ (Eq.~\eqref{chi1-def}).

By evaluating the time-averaged radiated power \cite{Novotny_book:2008} due to the SPP polarization at the second harmonic (Eq.~\eqref{dphi2-sol}) and the fundamental frequency (Eq.~\eqref{dphi_linsl}), we introduce their ratio $P(2\omega)/P(\omega) = 16 |\chi^{(2)}E^2_\texttt{in}|^2/|\chi^{(1)} E_\texttt{in}|^2$ quantifying the SHG efficiency.\footnote{Here, numerical prefactor of $16$ reflects scaling of the radiated power with emission frequency as $P(\omega)\sim\omega^4$.} With the help of Eq.~\eqref{chi2-def}, this expression simplifies to the form 
\begin{eqnarray}
\label{I2toI1}
\frac{P(2\omega)}{P(\omega)} = 16 {\cal N}_o\alpha^2 \frac{|\chi^{(1)}(\omega)E_\texttt{in}|^2}{(2\omega-\omega_o)^2+\gamma_{sp}^2},
\end{eqnarray}
explicitly indicating quadratic scaling of the SHG efficiency both with $\chi^{(1)}$ and the anharmonic coupling $\sqrt{{\cal N}_o}\alpha$.

\subsection{Numerical analysis}
\label{Sec:TC-numerics}

We preform the analysis by calculating the polariton dispersion (Eq.~\eqref{Lmbd_pm-def} and \eqref{plrt-poles}) and related absorption/emission spectra determined by $\text{Im}\left[\chi^{(1)}(\omega)\right]$ (Eq.~\eqref{chi1-def}), and the SHG efficiency (Eq.~\eqref{I2toI1}) depending on SPP-QE coupling strength ${\cal N}_o\lambda^2/\omega_o^2$. Here, and below  all the rates/frequencies are represented in units of $\omega_o$. The relaxation rates and incident field Rabi frequency are parameterized as follows  $\gamma_{sp}/\omega_o=7\times 10^{-3}$, $\gamma_\uparrow/\omega_o=7\times 10^{-3}$,  $\gamma_\phi/\omega_o=5\times 10^{-3}$ and $\Omega_\texttt{in}/\omega_o=4\times 10^{-3}$. The incoherent pumping rate is fixed at $\gamma_\uparrow=1.1\gamma_\downarrow$ corresponding to the population inversion corresponding to $d_o=0.05$. For these parameters, the critical coupling value is ${\cal N}_o\lambda_c^2/\omega_o^2=1.9\times 10^{-3}$. These parameters are similar to those adopted in the simulations discussed in Sec.~\ref{Sec:MaxSim}. Finally, we introduce two limiting regimes of weak, $\sqrt{{\cal N}_o}\alpha/\omega_o=10^{-3}$, and strong, $\sqrt{{\cal N}_o}\alpha/\omega_o=0.2$, anharmonicity and compare them below. 

Figure~\ref{Fig:Plr}a shows polariton frequency (cyan curve) and dephasing rate (red curves) as a function of the coupling strength between QEs and SPP calculated for the weak anharmonicity. The  dispersion follows the trends discussed in Sec.~\eqref{Sec:DispEffc}: Below strong coupling threshold, the frequency is constant, $\omega_\pm=\omega_o$, and the dephasing rate $\gamma_\pm$ linearly falls to zero (see Eq.~\eqref{Lmbd_pm-triv}). Above the threshold the frequency does not show visible changes but the polariton dephasing splits into the lower and upper branches, as a result of real  square root in Eq.~\eqref{Lmbd_pm-def}. Associated linear spectra presented in panel~(b) of Fig.~\ref{Fig:Plr}, show rise of a single absorption peak below the threshold flipping into the narrow negative dip above the threshold, i.e., emission of amplified incident field typical for the lasing steady state. This peak is associated with the lower-polariton dephasing branch, i.e., $\gamma_{-}\approx 0$, shown in panel~(a). The upper-polariton dephasing branch contributes to the widths of a broad absorption feature around the emission dip. Maximum amplification occurs in the vicinity of the critical coupling value. Further increase in the coupling strengths reduces depth of the dip. Panel~(c) of Fig.~\ref{Fig:Plr} shows more than five orders of magnitude enhancement in the SHG efficiency facilitated by the amplification for the coupling value slightly about the lasing threshold (red curve). According to Eq.~\eqref{I2toI1}, such an enhancement rises from quadratic scaling of the linear susceptibility compensating weak anharmonic coupling.

For the adopted strong anharmonicity value, the polariton dispersion (Fig.~\ref{Fig:Plr}(d)) above the strong QE-SPP coupling threshold is quite different compared to the weak anharmonicity case (panel~(a)). Increase in the anharmonicity makes expression under the square root in Eq.~\eqref{Lmbd_pm-def} negative, resulting in its imaginary values and subsequently in the splitting of the polariton frequency (cyan curve). However, the dephasing rate stays degenerate as pointed out in Sec.~\ref{Sec:DispEffc}. In this case, the absorption spectra shown in Fig.~\ref{Fig:Plr}~(e) acquire the splitting showing up in the strong coupling regime. In contrast to the weak anharmonicity case, the peaks do not become negative dips which is the signature of the incident field absorption rather than amplification. The SHG efficiency shown in panel~(f) of Fig.~\ref{Fig:Plr}, demonstrates maximum three orders of magnitude enhancement (compare peak values of red and black curves) but the absolute value of the maximum efficiency (red curve) is comparable with that in the weak anharmonicity regime (red curve in panel~(c)). In this case, the efficiency gains additional increase due to its quadratic scaling with the anharmonicity parameter ${\cal N}_o\alpha^2$ (see Eq.~\eqref{I2toI1}).                

\begin{figure}[t]
\begin{center}
\epsfig{file=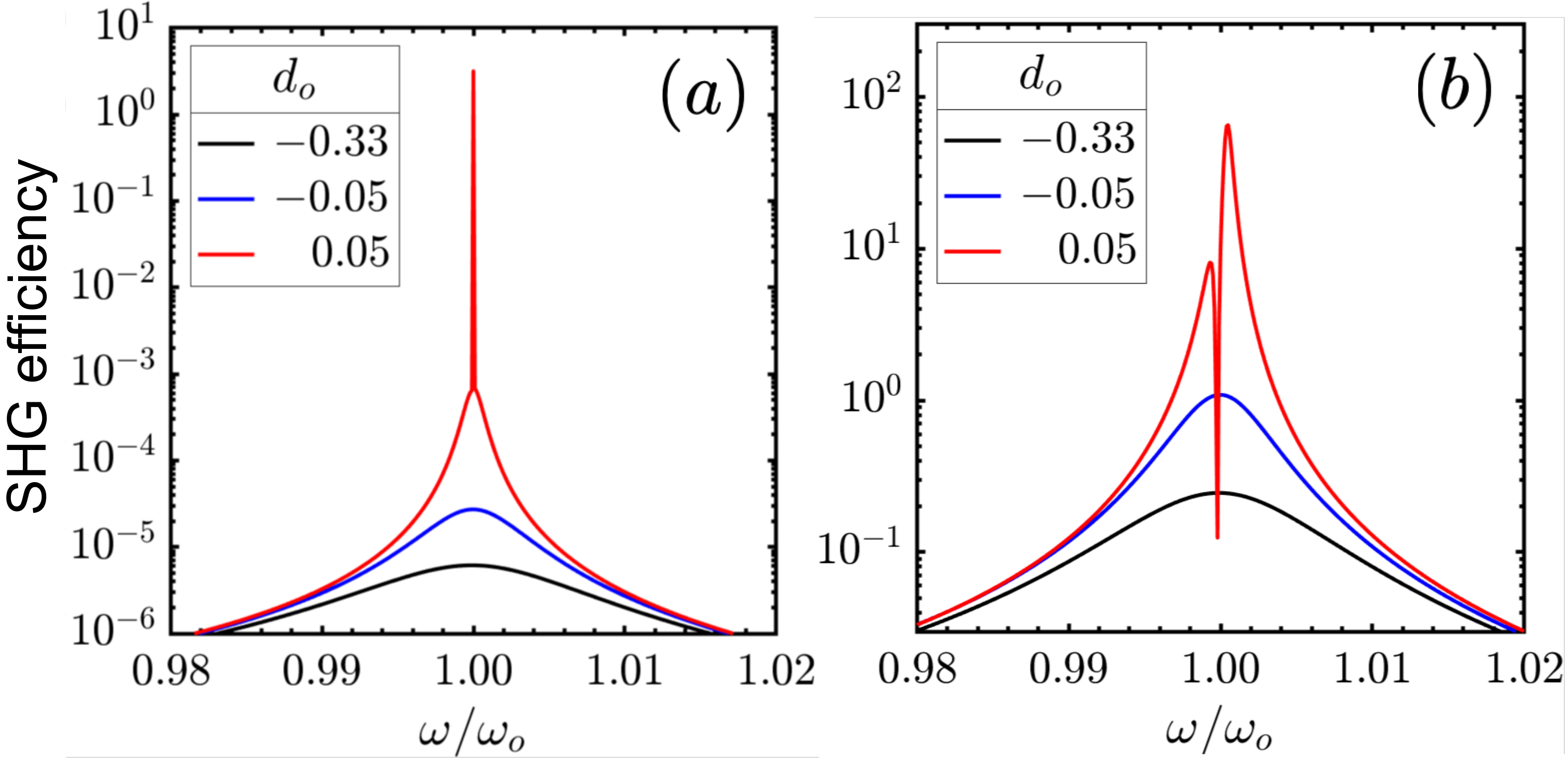,width=3.3in}
\end{center}
\caption{Comparison of the SHG efficiency for various values of the inversion parameter $d_o$ in the case of (a) weak, $\sqrt{{\cal N}_o}\alpha/\omega_o=10^{-3}$, and (b) strong, $\sqrt{{\cal N}_o}\alpha/\omega_o=0.2$, anharmonicity. The QE-SPP coupling strengths is fixed at ${\cal N}_o\lambda^2/\omega_o^2=2.0\times 10^{-3}$}. 
\label{Fig:I2/I1}
\end{figure}

Finally, we fix the coupling strength slightly above the critical coupling by setting ${\cal N}_o\lambda^2/\omega_o^2=2.0$ and vary the incoherent pumping rate below the inversion threshold to sample the states characterized by $d_o=-0.33$ ($\gamma_\uparrow=0.5\gamma_\downarrow$) and $d_o=-0.05$ ($\gamma_\uparrow=0.9\gamma_\downarrow$). Comparison of the SHG efficiencies calculated below inversion with that for the inverted QEs ($d_o=0.5$) are shown in panels~(a) and (b) of Fig.~\ref{Fig:I2/I1} for the weak and strong anharmonicities, respectively. As expected, the inverted system in the strong coupling regime shows tremendous enhancement of the SHG efficiency. However, comparison of Fig.~\ref{Fig:I2/I1}a with Fig.~\ref{Fig:Plr}c and Fig.~\ref{Fig:I2/I1}b with Fig.~\ref{Fig:Plr}f, clearly demonstrates that bringing the QEs to the population inversion but keeping the QE-SPP strength below the critical value does not result in significant SHG enhancement. Thus, to achieve high efficiency values, the system should be in the strong coupling regime. 

\section{SHG facilitated by two-dimensional lattice of metal nanopillars}
\label{Sec:MaxSim}

\begin{figure}[b]
\begin{center}
\epsfig{file=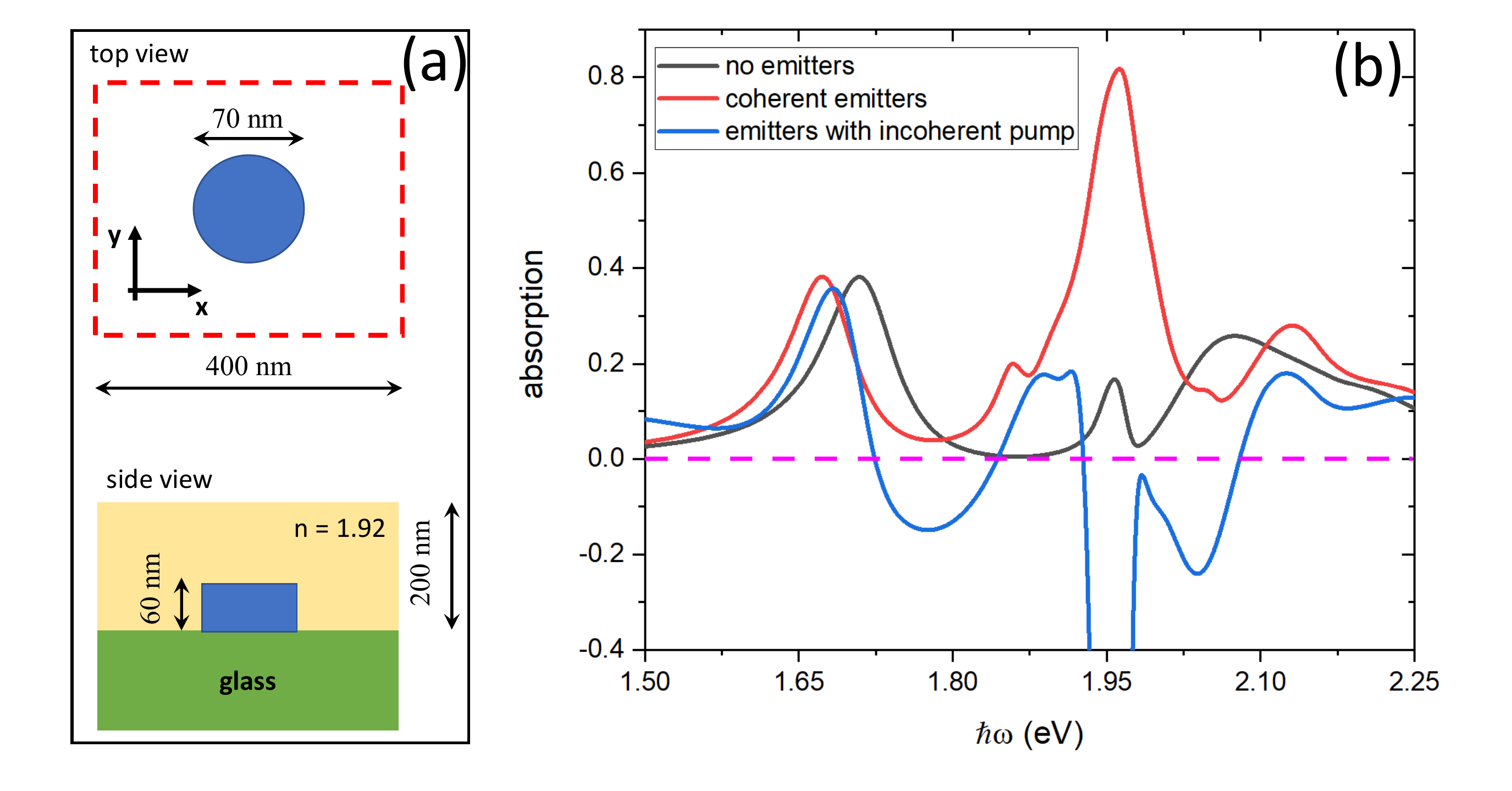,width=3.4in}
\end{center}
\caption{(a) Schematics of a repeat unit forming two-dimensional lattice. (b) Linear absorption spectrum of the plasmonic lattice calculated without QEs (black line), with QEs but no incoherent pump, $d_o = -1$ ($\gamma_\uparrow=0$), (red line), and in the inverted regime, $d_o = 0.048$ ($\gamma_\uparrow=1.1\gamma_\downarrow$) (blue line).}
\label{Fig:M1}
\end{figure}

\subsection{Computational model}
 
Here, we consider a periodic square lattice comprised of Ag nanopillars placed on top of semi-infinite glass substrate as schematically illustrated in Fig.~\ref{Fig:M1}a. The input side of the lattice is covered by a 200 nm thick dielectric. Such a setup has been shown to support both the localized plasmon (2.1 eV) and the surface lattice plasmon seen at $\hbar\omega_o=1.96$~eV in Fig.~\ref{Fig:M1}b. Combined with colloidal quantum dots this system was experimentally  demonstrated to exhibit lasing \cite{GuanACSNano:2020}. 

Similar to Sec.~\ref{Sec:anhTCM}, our computational semi-classical model treats QEs as two-level systems whose time-evolution is described by Bloch equations in the form
\begin{eqnarray}
\label{reqm-Pn}
&~&\left(\partial^2_t + 2\gamma_o\partial_t  +\omega^2_o \right){\bm P}_o 
= -\frac{2 \omega_o}{\hbar}\rho_o\bm \mu_o\left(\bm \mu_o\cdot{\bm E}\right)s_z,
\\\label{reqm-Pszn}
&~&\partial_t{s_z} +\gamma_t\left(s_z-\frac{d_o}{2}\right) =
        \frac{2}{\hbar\omega_o\rho_o}\left(\partial_t{\bm P}_o\cdot{\bm E}\right),
\end{eqnarray}
where ${\bm P}_o= \rho_o\bm\mu_o  s_x$ denotes the QE polarization with $s_x=$Re$[s_{-}]$; $\rho_o$ denotes QE density and $\bm\mu_o$ stands for the QE transition dipole. The relaxation rates in Eqs.~\eqref{reqm-Pn} and \eqref{reqm-Pszn} have the same definition as in Eqs.~\eqref{MF-sm} and \eqref{MF-sz}. However, in contrast to the Tavis-Cummings model, we explicitly introduce classical local electric field $\bm E$ interacting with QEs. This electric field is calculated by numerically solving Maxwell’s equations governing the dynamics of electromagnetic radiation including the coherent incident field. Optical response of the plasmonic lattice is modeled using nonlinear hydrodynamic model resulting in the following equation of motion for the metal polarization, ${\bm P(\bm r,t)}$,~\cite{DrobnyhJCP:2020}  
\begin{eqnarray}
\label{NSeq}
\partial_t^2{\bm P}&+&\gamma_e \partial_t {\bm P}  =
\\\nonumber&~&
\frac{e}{m_e^*}\left(n_0 e {\bm E} +\partial_t\bm P\times\bm B
    -{\bm E}\left(\bm\nabla\cdot{\bm P}\right)\right)
\\\nonumber&~&
    -\frac{1}{n_0e}\left(\left(\bm\nabla\cdot\partial_t{\bm P}\right)\partial_t{\bm P}
    +\left(\partial_t{\bm P}\cdot\bm\nabla\right)\partial_t{\bm P} \right).
\end{eqnarray}
Here, the plasma frequency is $\omega_p=\sqrt{n_0 e^2/\varepsilon_0 m_e^*}$, the electron gas damping parameter is set to $\gamma_e=0.14$~eV, and the number density of conduction electrons is $n_0 = 5.86\times 10^{28}$~m$^{-3}$. Note that in the linear regime Eq.~\eqref{NSeq} reduces to the conventional Drude model.

The QEs are tuned in resonance with the surface lattice plasmon mode at 1.96 eV and uniformly distributed in the dielectric surrounding the nanopillars. Coupled Maxwell-Bloch equations are propagated in time-domain using home-build codes utilizing finite-difference time-domain method and the weakly coupling method.~\cite{SukharevPRA:2011}  To speed up the calculations, we employ three-dimensional domain decomposition and split simulation domain into 1152 sub-domains, each carried by a single processor. With the spatial resolution of 1.5 nm and dimensions summarized in Fig.~\ref{Fig:M1}a the total number of grid points is $6.7\times 10^7$. To ensure numerical convergence, the time step is set at 2.5 attoseconds. Typical execution times of our codes vary between 20 minutes (for linear simulations) and up to 3 hours (for nonlinear simulations with the hydrodynamic model employed). Other parameters for emitters are as follows: number density is 10$^{26}$~m$^{-3}$, transition dipole moment is 10~Debye, the pure dephasing rate is $\gamma_\phi=10^{-2}$~eV, the non-radiative decay rate is $6\times 10^{-3}$~eV. The system is coherently driven by a plane wave linearly polarized along $x$-axis and propagating in $z$-direction normal to the lattice plane. Due to limited computational resources, the pulse excitation of 500~fs duration is used. For better numerical convergence of power spectra at high pump intensities, we implement the Blackman-Harris time envelope. All nonlinear results are obtained for the incident field amplitude of $E_\texttt{in}=2\times 10^{-2}$~V/nm.
 
\subsection{Simulation results}

\begin{figure}[b]
\begin{center}
\epsfig{file=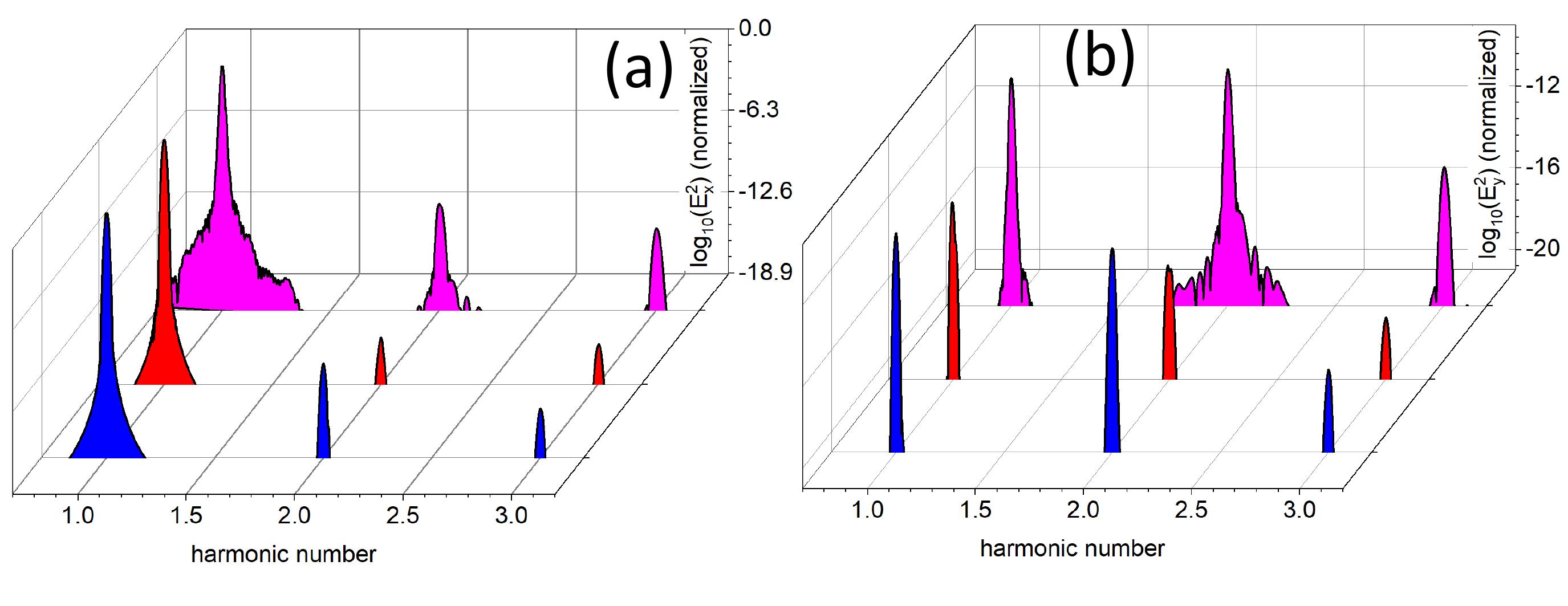,width=3.5in}
\end{center}
\caption{Nonlinear power spectra calculated for the plasmonic lattice without QE (blue), with QEs and no incoherent pump ($d_o = -1$) (red), and inverted QEs with $d_o = 0.048$ (magenta). Panel (a) shows signal polarized along the incident field polarization ($x$-axis) and panel (b) shows signal polarized perpendicular to the incident field ($y$-axis). Curves in both panels are normalized to the total signal value at the fundamental frequency.}
\label{Fig:M2}
\end{figure}

As a measure of the linear response of the plasmonic system we calculate absorption as a function of incident photon energy is shown in Fig.~\ref{Fig:M1}b. To simulate linear absorption we implement the short pulse method allowing to evaluate a wide frequency response in a single run.\cite{SukharevJPCM:2017} The system is excited by a short pulse and transmission, $T$, and reflection, $R$, are calculated. The absorption, $A$, is then evaluated according to the following expression: $A=1-T-R$. Simulations without QEs (black line) result in three distinct peaks nearly identical to those obtained in Ref. [\onlinecite{Guan_ASCnano:2020}]. The localized plasmon resonance is seen at 2.12 eV. The two other modes peaked at 1.67 eV and 1.96 eV are due to the surface lattice mode coupled to the guiding mode inside the dielectric covering nanopillars.

In the absence of incoherent pump ($\gamma_\uparrow=0$), the spectrum (red line) exhibits a large peak near QE transition frequency indicating response of strongly coupled QE-SPP states. Turing on the incoherent pump and setting the pumping rate to $\gamma_\uparrow=1.1\gamma_\downarrow$, we reach the inversion regime characterized by $d_o= 0.048$. In this regime the system acquires polariton gain resulting in the negative absorption (i.e., amplification) at the QE transition frequency (blue line). Interestingly due to strong coupling between QEs and SPP and significant material dispersion, the gain is also showing up at the nearby plasmon resonances (1.7~eV and 2.0~eV). Noteworthy that reduction in the incoherent pumping rate to the range of $0<d_o< 0.48$ destroys the gain. For example, a weak inversion with $d_o=0.020$ gives an absorption spectrum nearly equivalent to that with no pump at all. This can be understood by looking at Eq.~\eqref{lmbd_c-def} according to which and for $d_o\ll 1$, the critical parameter determining strong QE-SPP coupling regime scales as ${\cal N}_o\lambda_c^2/\sim 1/d_o$. The threshold quickly rises as $d_o$ approaches to zero. Accordingly for $0<d_o<0.048$ the systems gets into no amplification regime. Exact determination of the pumping threshold resulting in the gain effect is complicated by finite propagation time of the coupled Maxwell-Bloch equations. Even in the presence of gain, it may not be enough time to develop detectable increase in the signal to see the amplification effect. Due to complexity of the problem and thus long execution times the total propagation time is limited. Thus, for the analysis below we adopt the smallest inversion parameter, $d_o=0.048$, allowing for observable gain associated with 500~fs of propagation time.

Due to in-plane symmetry of the array, the second harmonic is generated primarily due to breaking up-down symmetry in $z$-direction, i.e. the pre-defined directionality of the $k$-vector of the pump results in SHG. It should be noted that the efficiency of SHG can, of course, be greatly enhanced by properly changing the shape of the nanoparticles (e.g., replacing nanopillars with L-shaped particles). For the adopted geometry with in-plane mirror symmetry and incident field polarized along $x$-axis, the diagonal element of the second order susceptibility tensor is $\chi^{(2)}_{xxx}=0$ and non-vanishing components  $\{\chi^{(2)}_{xyy},  \chi^{(2)}_{xxy}=\chi^{(2)}_{xyx} \}$ make the $y$-component of lattice polarization the primary contributor to the SHG. Fig.~\ref{Fig:M2} examines three scenarios: (blue) the array without emitters, (red) the array with emitters and no incoherent pumping ($d_o = -1$), and (magenta) the array of incoherently pumped QEs above inversion ($d_o = 0.048$). For all these cases, the incident field is driving the system at QE transition frequency. Here, we calculate separately power spectra polarized parallel ($x$-axis) to the incident driving field, Fig.~\ref{Fig:M2}a, and perpendicular ($y$-axis) to it, Fig.~\ref{Fig:M2}b. In the plot, the 2nd and 3rd harmonics are clearly observed  for both polarizations. 

We note that the $y$-polarized signal exhibits significantly stronger SHG (relative to the signal at the fundamental frequency) due to the symmetry considerations above. Once QEs are added, the SHG seen in the $y$-polarized signal is slightly reduced. This is primarily due to the fact that QEs do not generate 2nd harmonic directly and act as absorbers at the fundamental frequency. However, when conditions for the polariton  gain are met, the picture drastically changes. The $y$-polarized signal at the second-harmonic exceeds that at the fundamental frequency  as a result of strong local field enhancement. We note that this effect is predominantly observed for the $y$-polarized signal. This is again due to the in-plane symmetry and the fact that emission by the QEs is isotropic thus leading to the SHG enhancement only in the direction perpendicular to the incident field polarization. It is important to emphasize that the case with gain is not trivial as emitters do not directly generate SHG nor do they have pre-set coherences which would lead to the emission. The coherences are induced by the local interaction with corresponding plasmon mode of the lattice, which in turn enhances stimulated emission and this in turn leads to further local field enhancement significantly altering overall signal.

Choosing QEs resonant to the corresponding surface lattice plasmon mode may seem like an obvious choice to enhance SHG. However, due to strong coupling between QEs and SPP and because of high complexity of the parameter landscape that could significantly alter SHG, we scan the driving field frequency and explore the  SHG efficiency. Similarity to Sec.~\ref{Sec:DispEffc}, the latter is defined as the ratio of  radiated powers at the second harmonic and the fundamental frequency. When calculating the SHG efficiency, one needs to carefully take into account the material dispersion as transmission and reflection significantly varies with the driving field frequency.

\begin{figure}[b]
\begin{center}
\epsfig{file=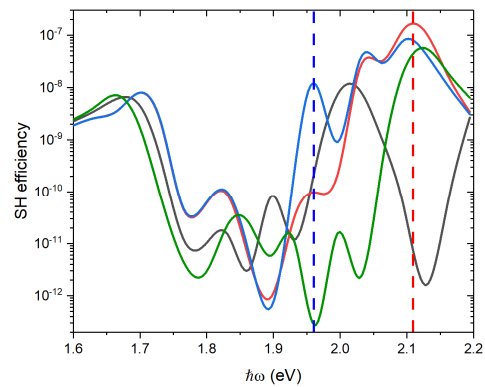,width=2.8in}
\end{center}
\caption{SHG efficiency as a function of the incident photon energy. Black and green lines show data with QEs with no incoherent pumping ($d_o = -1$) and transition frequency of 2.12 eV (black) and 1.96 eV (green). Blue line is for the QEs with $d_o = 0.048$ and the transition frequency of 1.96 eV. Red line shows results for the QEs with $d_o = 0.048$ and the transition frequency of 2.12 eV. Two vertical dashed lines indicate corresponding QEs' transition frequencies.
}
\label{Fig:M3}
\end{figure}

Fig.~\ref{Fig:M3} presents main results of this section. First, we consider no incoherent pumping of QEs, $d_o=-1$, (black and green lines). The SHG efficiency varies substantially over a wide range exhibiting strong enhancement near 2~eV and 2.12~eV (reaching value of 6$\times$10$^{-8}$), which corresponds to the localized plasmon resonance (see Fig.~\ref{Fig:M1}b). This is not surprising as the localized plasmon is characterized by a strong local field enhancement (which in turn leads to high damping since the mode is partially localized inside the metal). When QEs are resonant with the lattice plasmon mode at 1.96 eV, the SHG efficiency (blue line) displays similar complex behavior and peaks at the transition frequency. Additionally and most surprisingly, we observe another broad peak at 2.12~eV that is noticeably different from the QE resonance and at which the SHG efficiency is 10 times higher compared to the value at the driving field frequency. When QE transition frequency changed to 2.12~eV (red line), the SHG efficiency is slightly increased at the resonance compared to the off-resonant case (blue line). As expected, the peak at 1.96~eV is no longer seen but the enhancement of the SHG efficiency at 2.12~eV is not as strong as it is observed for 1.96~eV case. The latter raises the efficiency by 2 orders of magnitude compared to the case with no incoherent pump. When QEs are strongly coupled to the localized surface plasmon mode, our simulations demonstrate that the SHG efficiency is significantly increased. This is attributed to the local field enhancement near the surface of metal known to be higher for localized plasmon modes compared to the Bragg plasmons and lattice modes.

Comparing absolute values of SHG efficiency calculated using the anharmonic Tavis-Cummings model (Figs.~\ref{Fig:Plr}c,f and Fig.~\ref{Fig:I2/I1}) with full scale three-dimensional simulations (Fig.~\ref{Fig:M3}), it is important to emphasize several major differences between these models. First and foremost, the numerical simulations take into account multiple channels that lead to losses thus noticeably diminishing the SHG efficiency. Such channels include radiation to the far-field and spatially dependent local field distribution, which both significantly lower the coupling between QEs and corresponding resonant plasmon mode. Secondly, the anharmonicity parameter used in the model is most likely overestimated compared to the effective second order susceptibility, which directly comes from our hydrodynamic approach. The latter depends on the geometry used and as we pointed out the symmetry of square lattice along with a shape of each nanoparticle will affect the efficiency of SHG as well. Having non-centrosymmetric nanoparticles (L-shaped particles, nanocrescents, etc.) and arranging them in hexagonal lattices will further increase the anharmonicity and thus the SHG efficiency. 

\section{Conclusions}
\label{Sec:Concl}
We have considered plasmonic nanostructures interacting with two-level QEs under incoherent energy pump and demonstrated that in the strong coupling regime such a system exhibits noticeable enhancement of the SHG. The driven-dissipative Tavis-Cummings model accounting for the  anharmonic SPP, provides us with an insight into the mechanisms behind polariton gain and their interplay with the SPP anharmonicity. As demonstrated this results in strong SHG efficiency enhancement. However, this minimal model provides an upper boundary estimate for the SHG efficiency. 

To examine more realistic scenarios, we performed three-dimensional numerical simulations of the SHG by a square lattice of Ag nanopillars coupled to QEs. The numerical model employed is based on the fully vectorial nonperturbative nonlinear hydrodynamic model for conduction electrons coupled to Maxwell-Bloch equations for QEs. The simulations support the idea of the gain enhanced second harmonic generation showing orders of magnitude increase when inverted QEs are tuned in resonance with the lattice plasmon mode and brought to the strong coupling regime. By varying the driving field frequency and tuning QEs to a localized plasmon mode, we demonstrate further enhancement of the SHG efficiency facilitated by strong local electric fields. The incident light polarization dependence of the second harmonic generation is examined and related to the symmetries of participating plasmon modes.

Finally, we point out that our local anharmonic Tavis-Cummings model provides rather a qualitative insight into the results of the simulations based on the semiclassical  non-perturbative and non-local Maxwell-Bloch-hydrodynamic model. Establishing quantitative relationship between these models would require extraction of properly averaged $\chi^{(1)}(\omega)$ and $\chi^{(2)}(\omega)$ from the semiclassical calculations and their subsequent fitting with Eqs.~\eqref{chi1-def} and \eqref{chi2-def}. This is a subject of a separate study focused on qunatization of anharmonic plasmonic cavities extending beyond the scope of this report.

\section{Author's contributions}
\label{Sec:contributions}
All authors contributed equally to this work.

\section{Acknowledgements}
\label{Sec:Acknowledgements}

This work was performed, in part, at the Center for Integrated Nanotechnologies, an Office of Science User Facility operated for the U.S. Department of Energy (DOE) Office of Science by Los Alamos National Laboratory (Contract 89233218CNA000001) and Sandia National Laboratories (Contract DE-NA-0003525). M.S. is grateful for the financial support by Air Force Office of Scientific Research under Grant No. FA9550-19-1-0009.

\section{Data availability}
\label{Sec:data}
The data that support the findings of this study are available from the corresponding author upon reasonable request.


%

\end{document}